\documentclass{article}
\usepackage{amssymb,epsf}
\usepackage{graphicx}
\usepackage[dvips]{color}
\title{Random percolation as a gauge theory} 
\author{F.~Gliozzi$^a$, S.~Lottini$^a$, M.~Panero$^b$ and A.~Rago$^c$}
\newcommand{\eq}{\begin{equation}}
\newcommand{\en}{\end{equation}}
\newcommand{\ear}{\begin{eqnarray}}
\newcommand{\rae}{\end{eqnarray}}
\newcommand{\Z}{\mathbb{Z}}
\newcommand{\bra}{\langle}
\newcommand{\ket}{\rangle}

\definecolor{M_Beige}         {rgb}{0.96 , 0.96 , 0.86}

\definecolor{M_Brown}         {rgb}{0.65 , 0.16 , 0.16}

\definecolor{M_Gold}          {rgb}{1.00 , 0.84 , 0.00}

\definecolor{M_LemonChiffon}  {rgb}{1.00 , 0.98 , 0.80}

\definecolor{M_Orange}        {rgb}{1.00 , 0.60 , 0.00}

\definecolor{M_Pink}          {rgb}{1.00 , 0.75 , 0.80}

\definecolor{M_Violet}        {rgb}{0.93 , 0.51 , 0.93}

\newcommand{\resection}[1]{\setcounter{equation}{0}\section{#1}}

\begin{document}
\maketitle
\noindent
 $^a$ Dipartimento di Fisica Teorica, Universit\`a di Torino and\\ INFN,
sezione di Torino, via P.~Giuria, 1, I-10125 Torino, Italy.\\
$^b$ Dublin Institute for Advanced Studies,\\
 10 Burlington Road, Dublin 4, Ireland.\\
$^c$ Dipartimento di Fisica, Universit\`a di Milano and\\
INFN, Sezione di Milano, Via Celoria, 16, I-20133 Milano, Italy.
\vskip0.2cm
\begin{tabular}{rl}
e-mail:&gliozzi@to.infn.it lottini@to.infn.it\\
&panero@stp.dias.ie antonio.rago@mi.infn.it
\end{tabular}
\vskip0.2cm
\begin{minipage}[0cm]{0.95\textwidth}
\vspace{-16cm}
\hfill DIAS-STP-05-02

\hfill IFUM-823-FT
\end{minipage}
\begin{abstract}
Three-dimensional bond or site percolation theory on a lattice can be 
interpreted as a gauge  theory in which the 
Wilson loops are  viewed as  counters of topological linking with
 random clusters. Beyond the percolation threshold large Wilson
loops decay with an area law and show the universal shape effects 
due to  flux tube quantum fluctuations like in ordinary confining gauge 
theories. 
Wilson loop correlators define a non-trivial spectrum of physical states 
of increasing mass and spin, like the glueballs of ordinary gauge theory.
The crumbling of the percolating cluster when the length  of one periodic 
 direction decreases below a critical threshold accounts for the 
finite temperature deconfinement, which
belongs to 2-D percolation universality class.

\end{abstract}

\resection{Introduction}
Percolation is a purely geometrical phenomenon which in many respects 
resembles a continuous thermal phase transition. The theoretical description of
the percolation processes is conventionally given in terms of the cluster sizes
\cite{stau}, and most of the universal scalings deal with size distribution 
of clusters. The point of view which is taken in this paper is different. 
We focus on  topological entanglement of random clusters and use
it to  describe how  percolation  theory in three dimensions can be viewed 
as a full-fledged gauge theory. 

Though the gauge group of the theory in question is trivial
(it
is the $q\to 1$ limit of the symmetric group $S_q$), the occurrence of 
a confining phase yields some new hints on the mechanisms of quark 
confinement of more general theories.  
  
Transcribing percolation in terms of gauge theory  has also some 
important  consequences for  percolation itself. In three-dimensional 
systems there are almost no exact results whatsoever, but in gauge theories 
we have  a number of well-verified conjectures that can be translated into 
the language of percolation.
In this way we shall, for instance, relate 
certain linking properties of the 
closed paths within a percolating cluster to the universal quantum 
fluctuations of the chromoelectric flux tube joining a quark pair in the 
confining phase of whatever gauge theory. 
 
An unsuspected property of random percolation which emerges from this new 
 viewpoint is that its  spectrum  should be 
composed by a (possibly infinite) tower of physical states of increasing mass 
and spin --- the glue-balls of the corresponding gauge theory. Their mass ratios
near the percolation point define a totally new set of critical amplitude 
ratios belonging to the universality class of  3-D random percolation.       

Another piece of useful information comes from considering  percolation in 
a slab which is infinite in two dimensions, but of finite length $\ell$ 
and periodic in the remaining direction. The associated gauge model 
describes a system at finite temperature $T=1/\ell$.  This  transition 
is accurately described  by the universality class 
of two-dimensional random percolation, but the corresponding deconfining  
temperature $T_c$ may be used to define a new critical amplitude of the 
three-dimensional system.

We test the universality of this new set of critical amplitudes by performing
large scale numerical experiments in three different kinds of lattices: the 
simple cubic (SC) with bond or site percolation and the body centered cubic 
lattice (BCC) with  bond percolation. The numerical 
implementation of these systems is straightforward in comparison with 
simulations of ordinary gauge models: no Markovian process is needed and 
there are neither
thermalization problems nor critical slowing down. 
Preliminary results have been presented in  \cite{Gliozzi:2003mb}.

The organisation of the paper is as follows. In the next Section we define 
a new class of observables of the percolation theory to be identified with 
the Wilson loops of the corresponding gauge theory. In Section~\ref{cut} 
we describe a method to study the linking properties of the random clusters 
which is at the heart of our analysis. Section~\ref{confin} is devoted to the 
comparison of confinement mechanisms and in Section~\ref {string} we 
extract the string tension and show the relevance  of the 
universal terms generated by quantum fluctuations of an underlying effective 
string.
In Section~\ref{spectrum} we study the plaquette correlators in order to 
find the low-lying states of the spectrum. Section~\ref{slab} 
is devoted to the transcription of the percolation 
on a slab into a gauge system at 
finite temperature and Section~\ref{universality} discusses the universality 
class of the finite temperature deconfinement transition. Finally in 
Section~\ref{conclusion} we draw some concluding remarks.

\resection {Observables}
\label{observa}
The most basic observables of any gauge theory are the Wilson loops. 
These are operators 
which assign to each pair $(C,\gamma)$  formed by an arbitrary gauge 
configuration $C$ and any closed path $\gamma$ of the space  a 
suitably defined complex number $W_\gamma(C)$. Their importance stems 
from the fact that they serve as order parameters for confinement: 
in the confining phase  the vacuum expectation value of large Wilson loops 
exhibits area law. 

One can  define  
similar observables in the framework of random percolation.
In this context the  configurations are generated simply by occupying 
each site or bond on an initially empty 3-D lattice $\Lambda$ with 
independent probability $p$. Two sites are considered neighbours if they share 
one bond. The resulting configuration is a graph $G$ drawn on the lattice, 
composed by the occupied bonds 
(bond percolation problem) or by the bonds joining occupied neighbour sites 
(site percolation problem). The connected components of $G$ are the 
{\sl clusters} of the configuration. We choose as $\gamma$'s the closed 
paths of the dual lattice $\tilde{\Lambda}$. The value $W_\gamma(G)$ measures 
the topological entanglement between $\gamma$ and $G$. More precisely we 
apply the following
rule 
\begin{enumerate}
\item $W_\gamma(G)=1 $ if no cluster of $G$ is  linked to $\gamma$;
\item $W_\gamma(G)=0$ otherwise.
\end{enumerate}         

This definition did not come out of the blue. Starting from the 
Fortuin-Kasteleyn random cluster  representation of the 3-D Ising model 
\cite{fk} combined with the Kramers-Wannier duality \cite{kw}, it
is easy to 
express the Wilson loop $W_\gamma$ belonging to a $\Z_2$ gauge system 
in terms of the winding numbers modulo 2 of the Fortuin-Kasteleyn clusters 
across the $\gamma$ loop   \cite{Gliozzi:1996fy}. Generalising this result to
 $q$-state Potts model one is led to the above rule in the case of non-integer $q$.
A recently developed algorithm could even allow to evaluate explicitly these 
quantities for  real $q>0$ \cite{Gliozzi:2002ub}.
We extend the above recipe to any percolating system, owing to the fact that 
bond percolation can be viewed as the $q\to1$ limit of the 
$q$-state  Potts model.

\begin{figure}
\begin{center}
\centering
\includegraphics[width=0.8\textwidth]{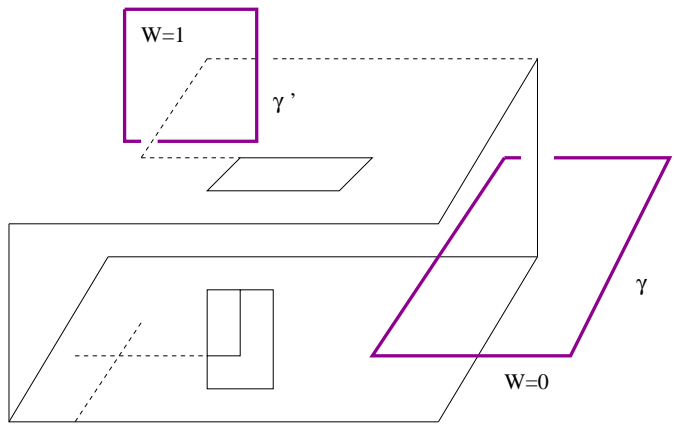} 
\caption{The closed thick lines $\gamma$ and $\gamma'$ represent
 Wilson loops. The dashed lines are bridges of the cluster. The
other solid lines are closed paths of the cluster. $\gamma$ is 
linked to the cluster, while $\gamma'$ is unlinked. }
\label{Figure:0}
\end{center}
\end{figure}

The removal of an occupied bond $b$ from a graph $G$ can lead 
to two different issues. If the number $c(G)$ of clusters  is kept invariant,
then $b$ is  necessarily a step of a closed path (or {\sl loop}) of $G$,  
whereas if $c(G)$ increases by
one,  $b$ cannot lie in a loop and is called a {\sl bridge} 
(see Fig.~\ref{Figure:0}). 
 Clearly only the former bonds contribute to $W_\gamma(G)$. If two graphs
$G$ and $G'$ have the same loops and differ only in the bridges, then they 
yield the same value of $W_\gamma$ for any $\gamma$. In more technical 
terms we may write the double implication
\eq
G\cup G'-G\cap G'\,\;\;{\rm is~a~tree}\;\;\;
\Leftrightarrow \;\;\;\,W_\gamma(G)=W_\gamma(G')~,
\forall\,\gamma\subset\tilde{\Lambda}~~,
\label{similgauge}
\en 
hence the transformation $G\to G'$ has some resemblance 
to a gauge transformation. 

Note that the connected correlator among occupied 
bonds is exactly zero by construction, but this is not an invariant quantity 
under the above  $G\to G'$ transformation. Cutting all bridges of $G$ yields
the maximal  invariant subset $B_G$ of $G$, made by bonds belonging to 
some loop. Of course we have $W_\gamma(G)=W_\gamma(B_G)$ for any $\gamma$. 
 The connected correlator among bonds \underline{belonging to $B_G$} 
is by no means trivial and is directly related to the connected correlator 
of the  plaquette, {\sl i.e.} the Wilson loop $W_\square$ associated to the 
smallest loop $\square\subset\tilde{\Lambda}$, because    
$W_\square=0$ if and only if the bond dual to $\square$ belongs to $B_G$.
In Section~\ref{spectrum} we shall use such a correlator to extract 
the low-lying mass spectrum of the theory.

\begin{figure}
\begin{center}
\centering
\includegraphics[width=0.8\textwidth]{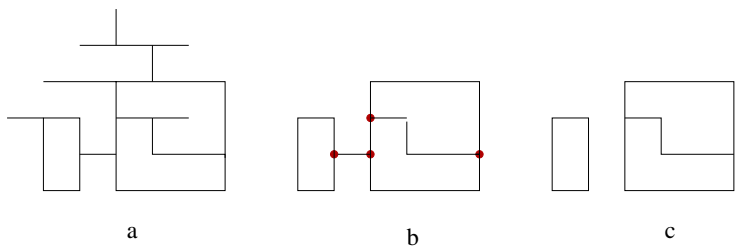} 
\caption{Cutting all bridges of a configuration}
\label{Figure:1}
\end{center}
\end{figure}

\resection{Cutting all bridges}
\label{cut}
In our approach the only bonds which play a role in the evaluation of the 
 observables defining the gauge theory are those belonging to loops.
Thus, once a new configuration $G$ is generated, we first get rid of all
 bridges. One way  of achieving this goal is the following.
\begin{enumerate}
\item Eliminate all the dangling ends (see Fig.~\ref{Figure:1}a and b).
At this stage the 
remaining graph is formed by loops and lines of bridges connecting them.
\item Build a reduced graph in which the only vertices are the lattice 
sites with more than two incident occupied bonds. The edges of the graph
are formed by lines of occupied bonds (Fig.~\ref{Figure:1}b).
\item Erase one edge at a time and apply each time a cluster reconstruction 
algorithm (for instance Hoshen-Kopelman \cite{hk}) in order to pick off the remaining 
bridges (Fig.~\ref{Figure:1}c).
\end{enumerate} 

\begin{figure}
\begin{center}
\centering
\includegraphics[width=0.8\textwidth]{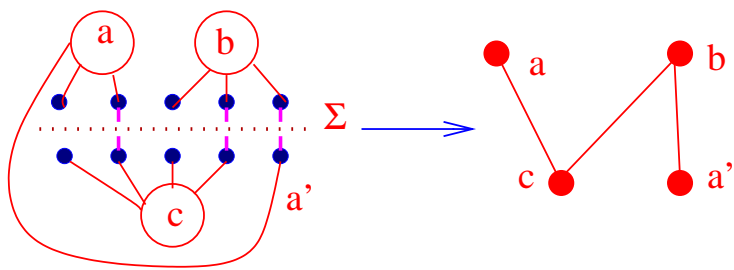} 
\caption{A 2-D sketch of the method used to evaluate the linking of
a configuration with a loop $\gamma$. The two lines of solid dots represent 
the sites on  either side of the layer of  bonds piercing the 
surface $\Sigma$ (dotted line). The vertical broken lines are the switched 
off bonds. The big open circles represent the clusters of the cut graph 
to which the sites of the layer belong. The cluster $a$ reaches both 
margins of the layer. }
\label{Figure:2}
\end{center}
\end{figure}

In order to check whether a planar loop $\gamma\subset \tilde{\Lambda}$ 
is linked to a configuration $G$, we first project out all bridges as 
discussed. Then we switch off the layer of occupied bonds which pierce the 
planar surface $\Sigma$ encircled by $\gamma$ and   rebuild
the cluster structure of the {\sl cut graph}. For a non-trivial linking 
there must be at 
least one cluster which reaches the layer on either side (see the graph on
 the left of  Fig.~\ref{Figure:2}). In such a case we build an auxiliary 
graph in  which the vertices represent the clusters of sites on either side 
of the layer; a cluster connecting sites of opposite sides is represented 
by two vertices, one for each side (see the graph on the right of 
Fig.~\ref{Figure:2}). We draw 
an edge between two clusters in the opposite sides of the layer if they are
 connected by switched off bonds. The configuration is 
truly linked to $\gamma$ if there is a path (at least) joining two vertices 
lying on opposite margins of the layer, but belonging to the same cluster 
of the cut graph (like the vertices $a$ and $a'$ of the figure).    

The whole procedure of cutting all bridges of a configuration and then 
evaluating its linking property with a set of Wilson loops is  
time demanding,  thus
a good implementation is mandatory. However the variance of the 
measured quantities turns out to be small, hence one can reach more precise 
results than the corresponding estimates in ordinary gauge theories. 

\resection{Confinement mechanisms}
\label{confin}
According to the recipe given in the previous Section,
the vacuum expectation value of the Wilson operator $W_\gamma$ is defined as
\eq
\bra W_\gamma\ket=\lim_{N\to\infty}\sum_iW_\gamma(G_i)/N=
\frac{{\rm number~of~config.~unlinked~to~}\gamma}
{\rm total~number~of~ configurations}
\label{W}
\en
  What is the functional form of this quantity for large loops? It depends 
on the value of the occupation probability $p$. If $p$ is below the 
percolation threshold $p_c$, then there are only finite-size clusters.
If the loop $\gamma$ has much a larger size than the clusters,  
then the configurations where $W_\gamma(G_i)=0$  necesarily have
some cluster located near the loop. The  number 
of these clusters grows linearly with  the perimeter 
$\vert\gamma\vert$ of $\gamma$  and produces 
the exponential decay   $\bra W_\gamma\ket\propto 
e^{-\alpha\vert\gamma\vert}$. We say that the theory is deconfined. On 
the contrary if $p>p_c$ the theory is confined. Confinement is expected 
to show up in an area law
  for the expectation value of large Wilson loops. Indeed in such a case 
there is an infinite cluster and the number of closed 
paths linked to $\gamma$ grows with the area $A$ of the minimal surface 
$\Sigma$ encircled by $\gamma$. These paths will pierce $\Sigma $ at points; 
 let $N=\alpha A$ be their mean number. Assuming these points to be 
randomly distributed on the surface, the probability of finding $k$ 
such 
points inside $\Sigma$ is binomial,
\eq
P_N(k)=\left(\matrix{ N\cr k\cr}\right) \alpha^k(1-\alpha)^{N-k}.
\label{bino}
\en    
Note that only the $k=0$ term contributes to the numerator of (\ref{W}), so 
the expectation value of the Wilson loop becomes
\eq
\bra W_\gamma\ket=(1-\alpha)^N=e^{-\sigma A}~,~~\sigma=-\alpha\log(1-\alpha)~.
\label{area}
\en
One thus apparently obtains an area law decay with string tension
 $\sigma$ for any Wilson loop, including those of small size. There is 
however a flaw in the argument;
 even if the configurations are obtained by populating each bond 
(or site) of the lattice \emph{independently} with a probability $p$, when 
all the bridges are erased it is no longer true that the remaining bonds 
are randomly distributed, as anticipated previously. In fact, since the 
interaction among the intersection points of $\Sigma$ is rather weak, we 
expect an area law only for large enough Wilson loops.

In  gauge theories two different confining mechanisms were proposed. 
One is based on the condensation of center vortices \cite{th}. 
These objects are string-like
 structures which are created by gauge transformations with a non-trivial 
homotopy associated to the center of the gauge group $
{\cal C}(G)$. In a 3-D lattice the center vortices are represented by a skein 
of loops forming an infinite  network  in the confining phase
\cite{Engelhardt:1999fd}. Each center vortex linked with a Wilson loop
contributes to it with a factor of $\zeta\in{\cal C} (G)$, thus each 
configuration contributes with a factor $\zeta^n$, where $n$ is the number 
of linked loops. Although at the end the overall effect is again a decay 
of $\bra W_\gamma\ket$ with an  
area law, such a mechanism is slightly different from the one we have 
described in pure percolation, where a single linked loop suffices to gives
a weight zero to the configuration. Another difference is that center vortices 
carry some conserved charge. For instance, if ${\cal C}(G)=\Z_N$
the   vortex flux is conserved modulo $N$, while  paths of random 
percolation do not carry any conserved charge and can intersect freely. 

The other confining mechanism is based on the old conjecture \cite{ma} that 
the vacuum behaves like a dual superconductor.  
  The key element of this picture is the monopole condensate which squeezes the
 gauge field generated  by a pair of  sources (quarks)
into a thin flux tube (the dual version of the Abrikosov vortex).  
This causes the Wilson loop to decay with an area law. One is led to 
conjecture that such a thin flux tube should vibrate as a free 
string \cite{lsw}. As a consequence, the expectation value of 
a rectangular Wilson loop of size $R\times T$ is expected to have the 
following  asymptotic functional form in the continuum limit \cite{ol}
  \eq
\bra W(R,T)\ket=C\,e^{-p(R+T)-\sigma RT}
\,\sqrt{\frac{\eta(i)\sqrt{R}}{\eta(iT/R)}}~,
\label{frees}
\en
where $C,p$ and $\sigma$ are functions of the coupling constant and $\eta$ 
is the Dedekind function
\eq
\eta(\tau)=q^{1/24}\prod_{n=1}^\infty(1-q^n)~, ~~ q=e^{2i\pi\tau}~.
\label{eta}
\en  
The factor under the square root accounts for the universal quantum 
contributions of the supposedly string-like  flux tube describing 
the  interaction between far-away sources.

Random percolation, lacking any non-trivial conserved charge, can hardly 
account for effects which play the role of magnetic monopole 
condensation. Notwithstanding this difficulty, 
we get indirect evidence of the formation of a vibrating string-like 
flux tube by measuring the universal shape effects it produces, as 
discussed in the next Section.

\resection{String tension}
\label{string}
\[
\begin{array}{cc}
\refstepcounter{figure}
\label{Figure:3}
\epsfxsize=.471\textwidth
\epsfbox{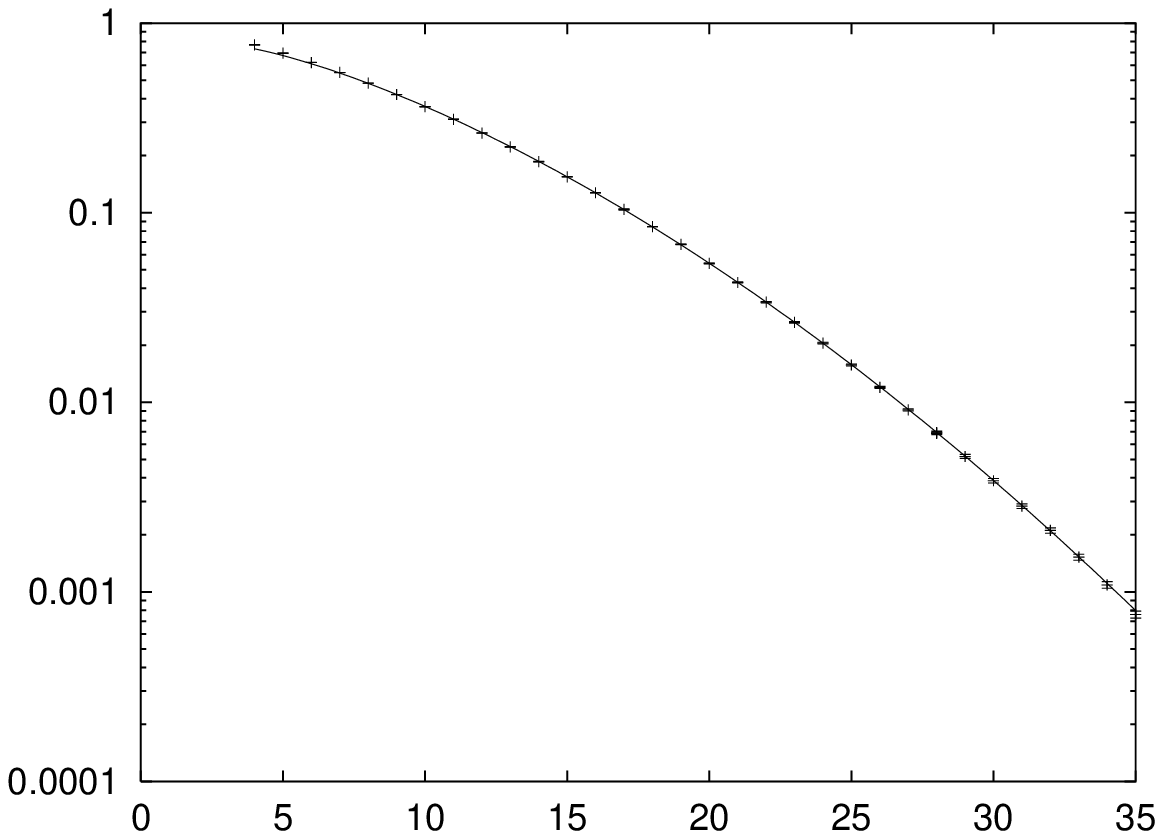}
&
\refstepcounter{figure}
\label{Figure:4}
\epsfxsize=.471\textwidth
\epsfbox{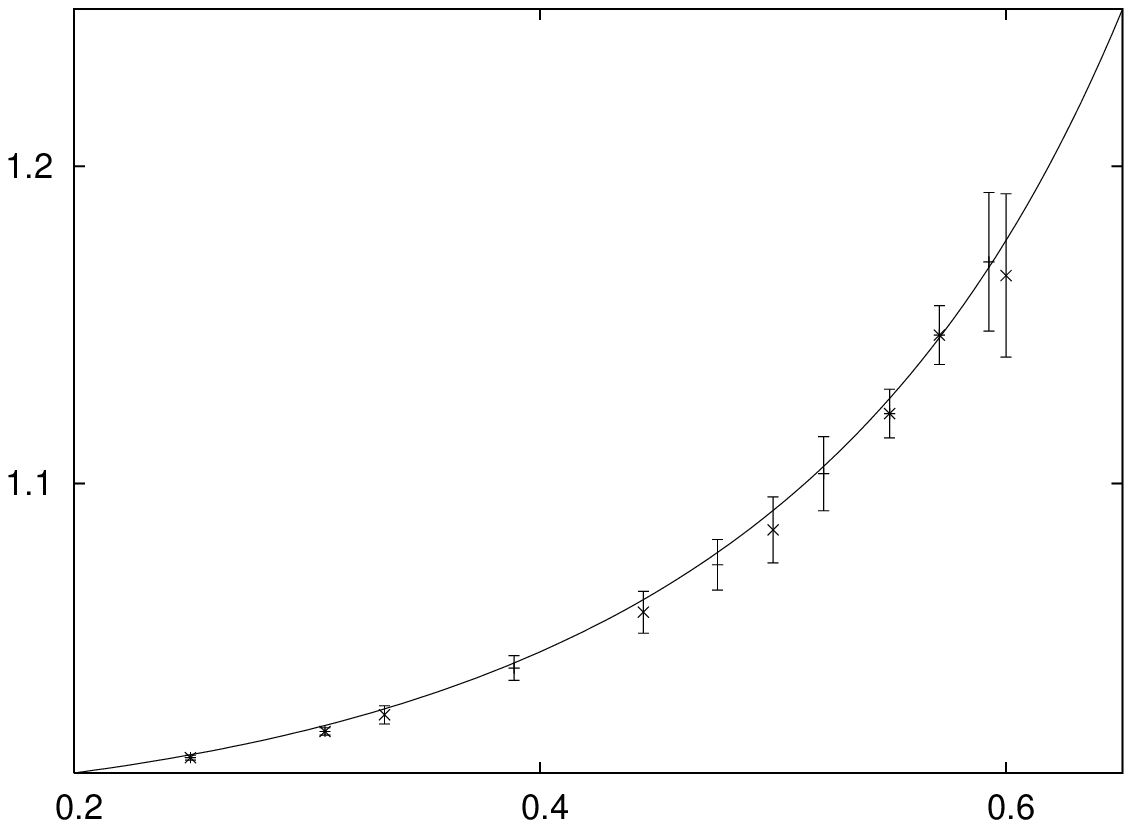}
\\
\parbox{.471\textwidth}{\small \raggedright
Figure~\ref{Figure:3}: Square Wilson loop as a function of $R$ for bond 
percolation in a cubic 
lattice of size $L^3=64^3$ at $p=0.26$~.} &
\parbox{.471\textwidth}
{\small \raggedright Figure~\ref{Figure:4}:
The universal shape effects of Eq. (\ref{qflu}) are 
compared with numerical data of Eq. (\ref{wrn}) for three different values of
 $p$.}
\end{array}
\]

We estimated the string tension $\sigma$ by fitting the mean values of the 
Wilson loops associated to squares of side $R$ to Eq.~(\ref{frees}), that in such a case becomes
\eq
W(R)=C\,R^{\frac14}\exp(-2p\,R-\sigma\,R^2)~~.
\label{wsq}
\en
Typically, in a lattice of size $L^3$ we considered  all the squares with 
$R\leq L/2$. The fits for not too small $R$
 are very good (see Fig. \ref{Figure:3}), nevertheless the parameters  
slightly depend on the value $R_{min}$ of the smallest square included 
in the fit.
Since  these formulae are expected to be valid only 
asymptotically for large values of $R$, we progressively eliminated the 
data of lower $R$ until stable parameters were obtained.

In order to check for presence of universal shape effects ascribed to the 
quantum fluctuations of the effective string,  we considered, as in 
\cite{Caselle:1996ii}, the quantity
\eq
{\cal R}(n,R)\equiv e^{-n^2\sigma}
\frac{W(R+n,R-n)}{W(R)}~~,
\label{wrn}
\en
which asymptotically ({\sl i.e.} large $R$ and $R-n$)
should be only a function of the ratio $t=\frac nR$, namely,
 \eq
{\cal R}(n,R)\to f(t)=\sqrt{\frac{\eta(i)\sqrt{1-t}}
{\eta(i\frac{1+t}{1-t})}}~~.
\label{qflu}
\en

\begin{figure}[htb]
\begin{center}
\centering
\includegraphics[width=0.9\textwidth]{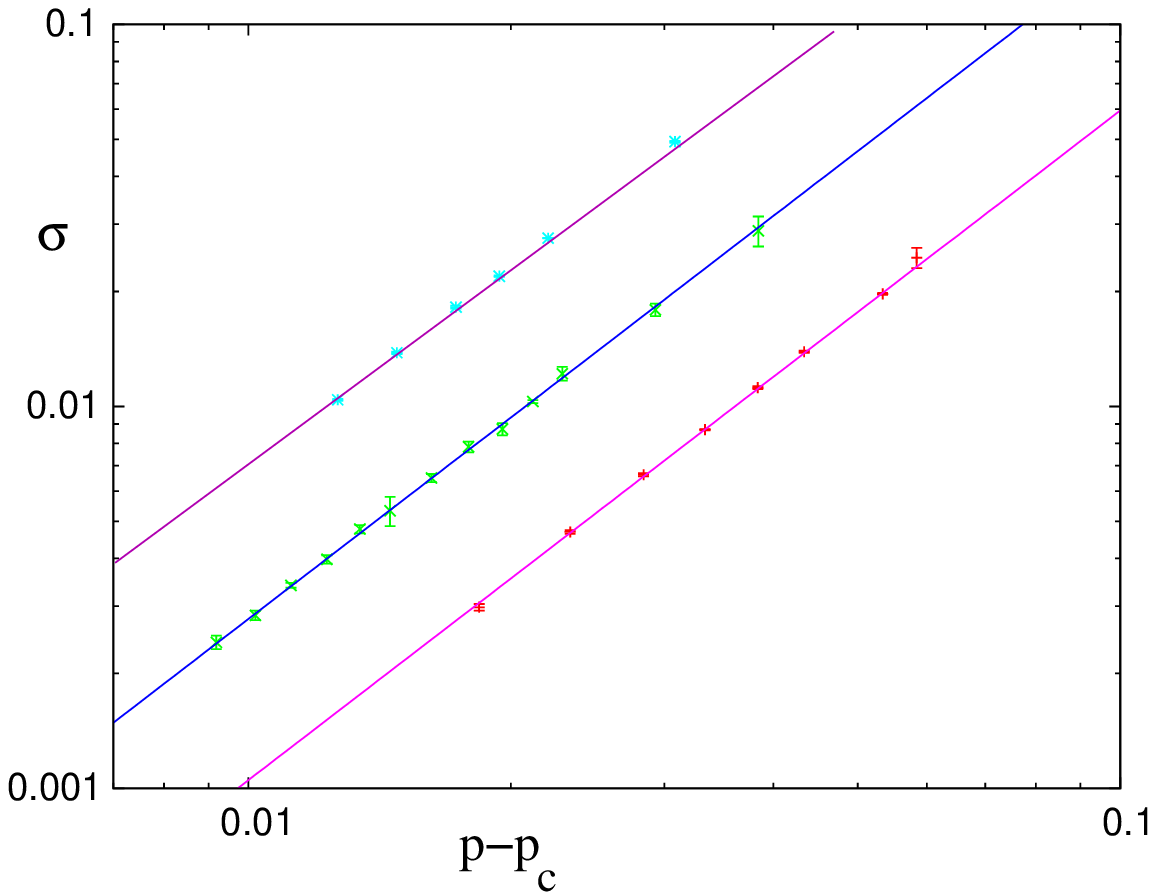} 
\caption{The string tension $\sigma$ as a function of $p-p_c$ for three 
different lattices: BCC bond (top line) SC bond (middle line) and SC site. 
The three parallel lines are one-parameter fits to Eq.~(\ref{sscaling}). The 
corresponding amplitudes are reported in Table~\ref{Table:1}.}
\label{Figure:5}
\end{center}
\end{figure}

Note that it does not contain any adjustable parameters. This function 
is plotted in Figure~\ref{Figure:4} and compared with the numerical data 
for three different values of $p$. The presence of the expected universal 
shape effects seems uncontroversial.

The string tension $\sigma$ is a physical quantity with the dimensions of an
inverse square length, hence it is expected to exhibit the following  power 
law behaviour sufficiently close to the percolation threshold 
\eq
\sigma=S\,(p-p_c)^{2\nu}~~,
\label{sscaling}
\en
where $\nu$ is the critical exponent of the correlation length in 3-D 
percolation. We used the value $\nu=0.8765(16)(2)$ of 
Ref.~\cite{Ballesteros:1998zm}\footnote {The first number between 
parenthesis is the statistical error, the second comes from the  
uncertainty in the scaling correction exponent $\omega$.\label{note}}.

\begin{table}[htb]
\begin{center}
\begin{tabular}{|c|c|c|c|}
\hline
Lattice&$p_c$&$S$&$\chi^2/d.o.f.$\\
\hline
SC site&0.3116081(7)(2) ~\cite{Ballesteros:1998zm}$~^{\ref{note}}$
&3.370(8)&1.15\\
\hline
SC bond& 0.2488126(5)~~\cite{lz}  &8.90(3)&0.30  \\
\hline
BCC bond& 0.1802875(10)~~\cite{lz}&22.07(2)&0.98\\
\hline
\end{tabular}
\end{center}
\caption{The amplitude of the string tension for three different lattices.  
Errors in parenthesis affect the last digits.}
\label{Table:1}
\end{table}

The percolation threshold $p_c$ depends on the lattice and on the kind of 
percolation (site or bond) which is 
studied. We checked Eq.~(\ref{sscaling}) in 
the percolating region of three different lattices: simple cubic 
(SC) with site or bond percolation and body-centered
cubic (BCC) with bond percolation. Precise estimates of 
$p_c$  are known  and are reported in Tab.~\ref{Table:1}. In the SC cases 
we used a lattice of size $L^3=64^3$, while in the BCC case we had
$L^3=55^3$.  In all cases the 
one-parameter fit to Eq.~(\ref{sscaling}) is very good and the scaling window 
seems rather wide (see Fig.~\ref{Figure:5}). The resulting amplitudes $S$ 
for the three lattices are reported in Tab.~\ref{Table:1}.  

Noteworthy, in order to extract the string tension $\sigma$ we have thus 
far assumed that the square Wilson loops obey the asymptotic form (\ref{wsq}), 
where the factor $R^{\frac14}$ accounts for the contribution of the string 
fluctuations.
If one neglected this factor and only took the area term into account, the 
$\chi^2$ test of the critical power law (\ref{sscaling}) would grow worse 
by one or two orders of magnitude, depending on the kind of lattice. 
We consider this fact another strong evidence of a vibrating, 
confining string.

\resection{Spectrum}
\label{spectrum}
As for standard  gauge theories, we expect that 
the confining phase of percolation possesses 
a rich spectrum of physical states with
increasing mass and spin, that we still call glueballs.
\begin{figure}[htb]
\begin{center}
\centering
\includegraphics[width=0.95\textwidth]{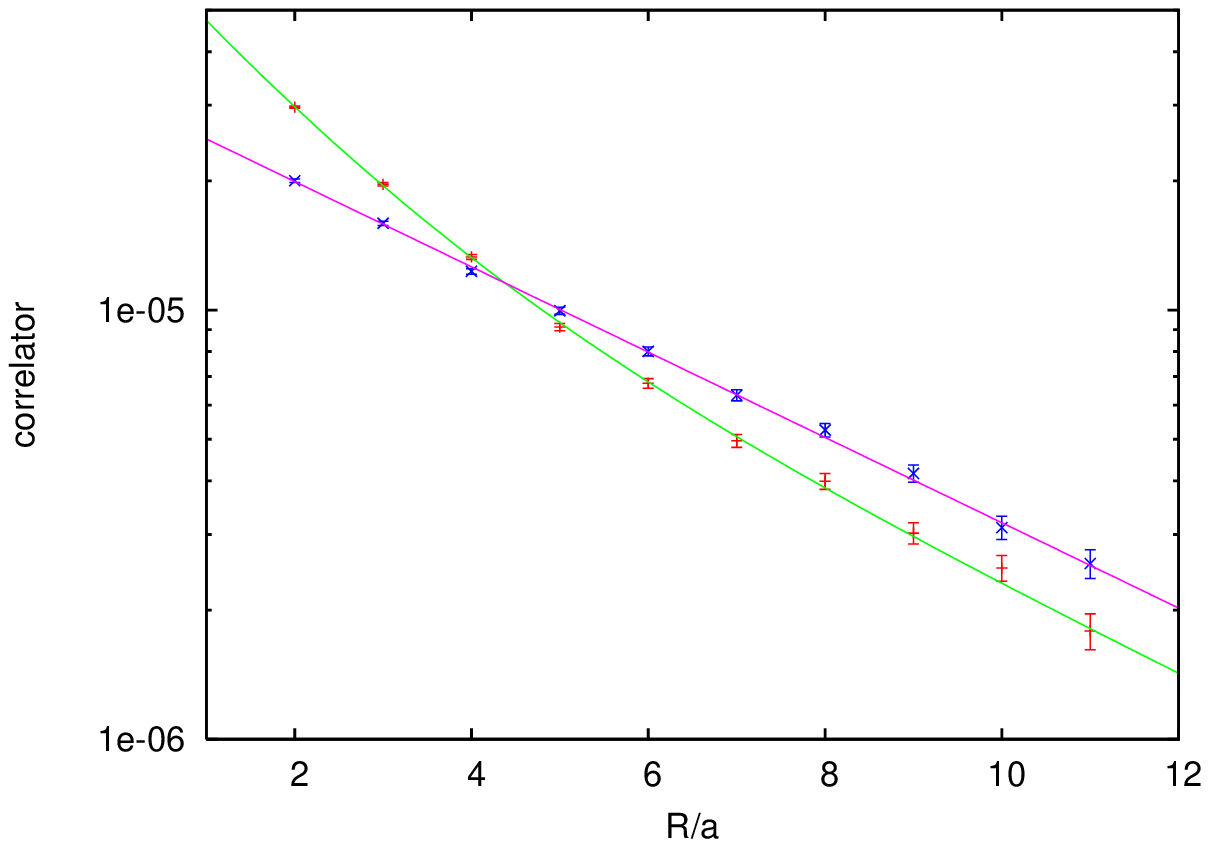} 
\caption{Comparison between the standard cluster correlator defined in Eq. 
 (\ref{clusc}) and the  connected plaquette-plaquette correlator  in a 
$32\times 32\times64$ cubic lattice at $p=0.260$. The straight line is an 
exponential fit to the cluster correlator data ($\times$). These data 
have been displaced downwards by two orders of magnitude for clarity.  
The other line is a two-exponential fit to the plaquette data (+). 
The semi-logarithmic plot makes it evident  that in  the latter case 
 a single exponential does not suffice.}
\label{Figure:6}
\end{center}
\end{figure}

\begin{figure}[htb]
\begin{center}
\centering
\includegraphics[width=0.8\textwidth]{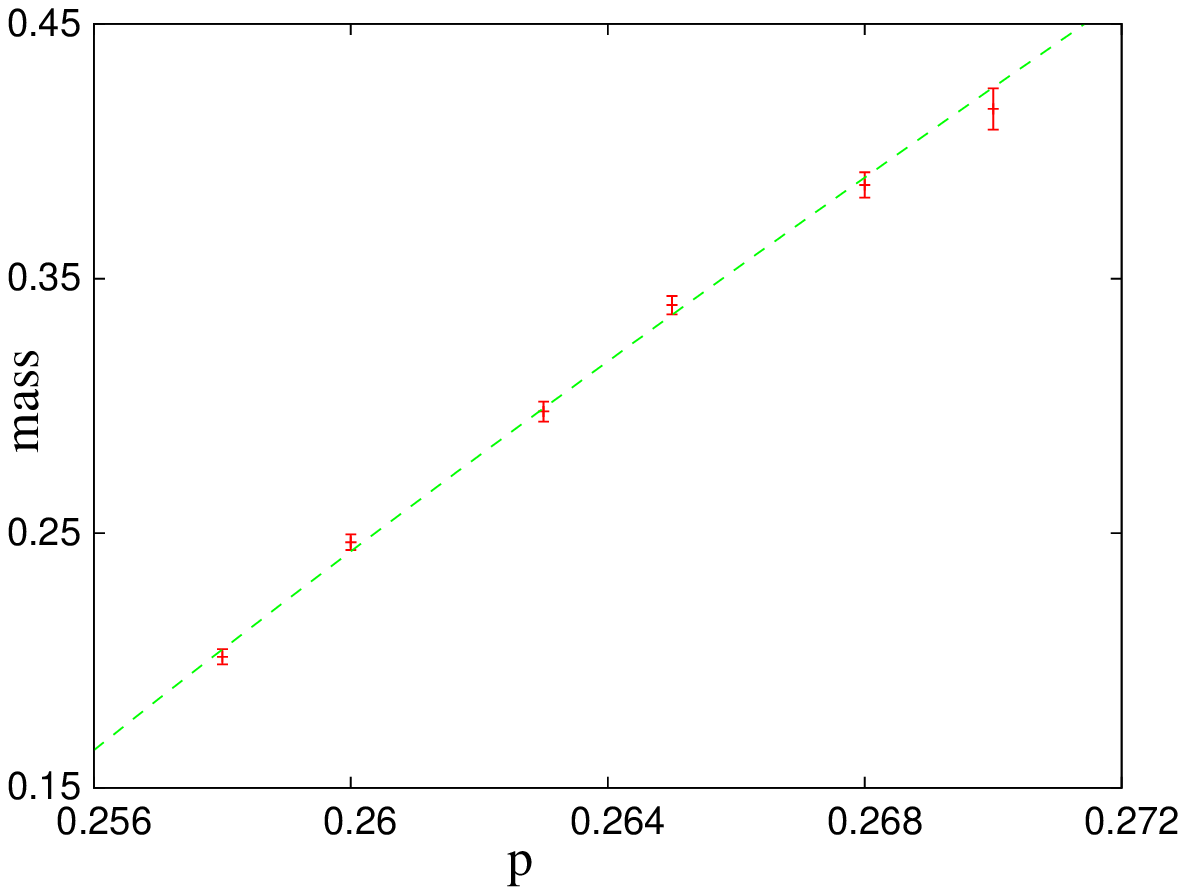} 
\caption{The mass of the lowest state as a function of $p$ in a simple 
cubic bond percolating lattice. The dashed curve is a one-parameter fit to 
Eq.~(\ref{masc}). }
\label{Figure:7}
\end{center}
\end{figure}

The basic method that goes into the computation of such mass spectrum is very 
simple. One first constructs a linear combination of Wilson loops on a 
fixed time slice of the three-dimensional lattice carrying the quantum 
numbers of the state one wishes to investigate. One then builds zero 
momentum operators by summing such a linear combination  over the entire 
spatial lattice. The simplest example of a zero momentum operator coupling to 
the spin $0^+$ states is given by
\eq
\Phi^{0+}(t)=\sum_{x,y}\left[W_{\Box\, 1}(x,y,t)+W_{\Box\, 2}(x,y,t)\right]~,
\label{glu0}
\en
 where $W_{\Box \,j}(\vec{x})$ denotes an elementary plaquette variable with 
base at $\vec{x}=(x,y,t)$ and orthogonal to the $j$ coordinate axis. 
According to Section~\ref{observa}, in random percolation the plaquette variable 
is replaced by the dual link variable $\ell_j(\vec{x})$, defined as equal to 1 
if the corresponding bond belongs to the subset $B_G$ of loops and null 
otherwise. The low-lying mass spectrum can be extracted by studying 
the exponential decay of the connected correlator
$C(t)=\bra\Phi^{0+}(t)\Phi^{0+}(0)\ket- \bra\Phi^{0+}\ket^2$, which is 
expected to have the following asymptotic expansion 
\eq
C(t)=\sum_n c_n\,e^{-m_nt}~~,
\en  
 where $c_n$ denote positive constants and $m_n$ are the glueball masses.
An example of such a correlator in a SC bond lattice is reported in 
Fig.~\ref{Figure:6}, where it is
 evident that at least two different scalar states contribute to $C(t)$. 
The estimates of the lowest mass in the range $0.258\leq p\leq0.270$ fit 
well with the  scaling form (see Fig.~\ref{Figure:7}) 
\eq
m=M_0\,(p-p_c)^\nu~,
\label{masc}
\en
with 
\eq 
M_0=12.45\pm0.07~.
\label{mass0}
\en  

Also the first excited state seems to follow the same power 
law as expected, though the errors are rather big. 
Its mass $M_0'$  is about twice $M_0$.
 
Such a 
behaviour is very different from the one observed in the standard 
two-point correlation function of the percolation problem, defined as the 
probability $G(\vec{x},\vec{x}')$ that the sites $\vec{x}$ and $\vec{x}'$ 
are in the same cluster\footnote{ This quantity is directly related to 
the correlator in $q\to 1$ limit of the $q$-state Potts model, 
see for instance  
Ref.~\cite{car}, page 156.}. The corresponding connected,  zero-momentum, 
projection
\eq
{\cal C}(t)=\sum_{x,y}\left[G(0,\vec{x})-G(0,\infty)\right]~~,
\label{clusc}
\en
exhibits a single exponential behaviour with a mass term which coincides,
within the numerical accuracy, with that of the lowest energy state coupled to 
the plaquette operator (see Fig.~\ref{Figure:6}). This indicates that the 
standard cluster correlator couples only to the lowest glueball, while the new
 observables suggested by the gauge theory interpretation of the percolation
disclose a totally unexpected spectrum of physical states. 

The lightest spinning glueball  is the $2^+$ state. It can be observed  
in the exponential decay of the correlation function of the operator
\eq
\Phi^{2+}(t)=\sum_{x,y}\left[\ell_1(x,y,t)-\ell_2(x,y,t)\right]~.
\label{glu2}
\en
A difficulty  encountered in this case is that the signal is drowned 
within the statistical noise for  values of $t$
beyond three or four lattice spacings.  
In spite of this accuracy problem, one can still verify that  in a SC 
bond lattice  the scaling form 
(\ref{masc}) is approximately obeyed with  an amplitude 
\eq
M_{2^+}=80\,\pm10~.
\label{M2}
\en
In order to have more accurate results on the mass spectrum the basis of the 
operators should be enlarged to Wilson loops of different shapes, trying to 
enhance their overlap with the glueball states.

\resection{Deconfinement at finite temperature}
\label{slab}
In quantum field  theory  the concept of temperature is introduced by 
simply compactifying the Euclidean time direction and identifying the inverse 
temperature with the temporal extension of the space-time manifold. Lattice 
field theories at a temperature $T$ are formulated in a slab which is 
infinite in the spatial dimensions, but of finite length $\ell=1/T$ 
(in lattice spacing units) and 
periodic in the remaining temporal direction. 

In any confining gauge theory there is a critical temperature $T_c$ above 
which the system is deconfined in the sense that for $T\ge T_c$ 
the string tension $\sigma$ vanishes. In this Section we demonstrate that 
the same phenomenon also occurs in random percolation. In the latter case 
the deconfinement mechanism is particularly transparent, showing its purely 
geometric origin: Non-vanishing string tension requires an infinite, 
percolating, cluster. Shrinking the width of the slab reduces the number of 
possible percolating paths along the spatial directions until the infinite 
cluster crumbles away, yielding a vanishing $\sigma$.

To put it in different terms, note that  as the temperature varies from 
zero to infinity a three-dimensional system is gradually dominated by  
two-dimensional behaviour; in particular the percolation threshold  is a 
decreasing function of the space dimensions. For instance in the 
SC bond lattice at $T=0$ the percolation threshold is at 
$p_c(D=3)=0.2488...$ (see Table~\ref{Table:1}). At $T=\infty$ the system 
reduces to a square lattice, where $p_c(D=2)=\frac12$, hence heating a 
system which at zero temperature lies in the percolating phase 
with $p<\frac12$  inevitably undergoes a deconfinement transition at 
a finite temperature $T_c$.
  
\begin{figure}[htb]
\begin{center}
\centering
\includegraphics[width=0.9\textwidth]{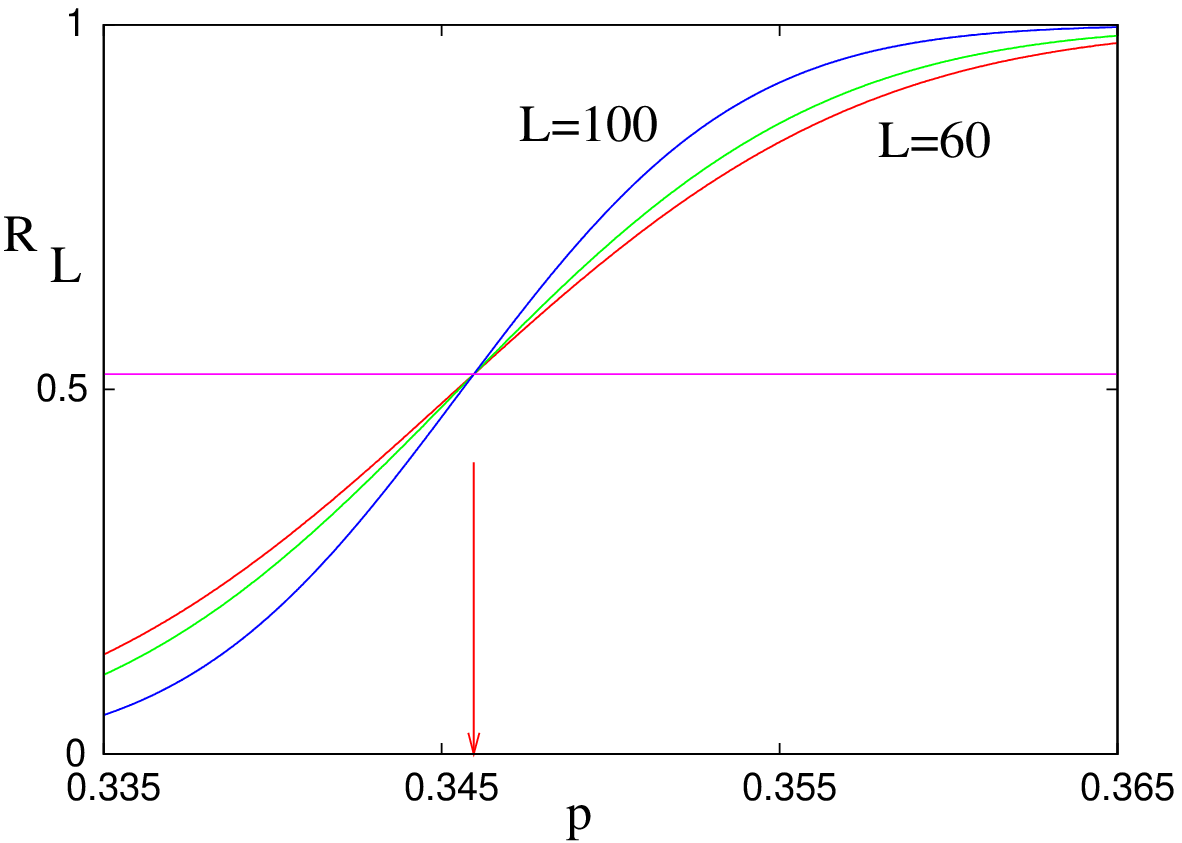} 
\caption{The wrapping probability $R_L$ in a simple cubic site percolating 
lattice for $\ell=7$ and $L=60,70,100$. The vertical arrow denotes the 
estimated value of $p_\ell$ at $L=\infty$ and the horizontal line indicates
the exact planar value of $R_\infty$.}
\label{Figure:8}
\end{center}
\end{figure}
In order to estimate $T_c$ in various  site or bond percolation lattices
we considered a slab of size $L_x \times L_y\times \ell$ with 
$L_x=L_y=L\gg\ell$ and periodic boundary conditions in all directions 
(see Fig.~\ref{Figure:11}).  
We calculated the probability 
$R_L(p)$ for a cluster to wrap around one of the large dimensions. Wrapping 
probabilities can be defined in different ways: wrapping around the $x$ 
direction, around the $y$ direction, around either direction, 
around both directions, etc. To be 
definite, we consider wrapping around the $x$ direction. For large 
$L$ it coincides  with the probability that the system percolates along $x$.

We evaluated $R_L(p)$ using a very efficient algorithm described by Newman 
and Ziff~\cite{nz}. For a bond percolation problem it consists of 
repeatedly adding a random bond to an initially empty lattice, identifying 
the clusters joined by the bond and merging
them if they are different. At 
each step one checks whether the touched cluster wraps around $x$ using a 
clever method described in Ref. \cite{mcl}. The process stops as soon 
as a wrapped cluster is detected. In this way one can evaluate the probability 
$Q_L(n)$ that a random configuration  with $n$ occupied bonds is wrapping 
around $L_x$ for any $n\leq N$, where $N$ is the total number of bonds of the 
lattice. This method may be adapted  to site percolation in a straightforward 
way. Then one simply finds the required 
quantity $R_L(p)$ for any value of $p$ 
by convolution with the binomial distribution
\eq
R_L(p)=\sum_n\left(\matrix{N\cr n\cr}\right) p^n(1-p)^{N-n}Q_L(n)~~.
\label{convolution}
\en  
Figure~\ref{Figure:8} shows some examples of $R_L(p)$ in the case of 
SC site percolating lattice. 
In the  $L\to\infty $ limit this wrapping probability becomes a step function. 
We have
\eq
R_\infty(p)=\cases{0 & for $p<p_\ell$\cr 1 &  for $p>p_\ell$\cr}~,
\en
where the threshold value $p=p_\ell$  depends on the type of lattice 
and on its width $\ell$.
When $L=\infty$  the slab system is equivalent to a 
2-D torus, where the wrapping probability at criticality has been 
calculated exactly by 
Pinson~\cite{pi} and is $R_\infty(p_\ell)=0.521058290...$ .  
We can use his result to measure the value of $p_\ell$ in a slab of width 
$\ell$  by finding the value of $p$ for which 
\eq
R_L(p)=R_\infty(p_\ell)~~.
\en

This method was first applied in the case of the Ising model in
 Ref.~\cite{Caselle:1995wn} (with the appropriate value of
 $R_\infty$, of course). These estimates in the case of 2-D percolation 
turn out to scale particularly well with the system size: Newman and Ziff 
argued that in planar lattices the leading order finite size correction  goes
like
\eq
p=p_\ell+c\,{L^{-2-1/\nu_2}}=p_\ell+ c\,{L^{-11/4}}~~,
\label{tscal}
\en
where $\nu_2=\frac43$ is the thermal exponent of 2-D random percolation.

 We checked it in the slab geometry finding good agreement for $L$ large 
enough, as Figure~\ref{Figure:9} shows in the case of a slab of width $\ell=6$.

We estimated in this way the threshold $p_\ell$ for seven different lattices.
The results are reported in Tab.~\ref{Table:2}. The slab widths 
(or equivalently the inverse temperatures) were chosen in such a way that 
the values of $p_\ell$ lie in the scaling region of the string tension 
$\sigma$, as determined in Section~\ref{string}. In this 
manner we  were able to evaluate  also the ratio $T_c/\sqrt{\sigma}$ 
using the amplitudes 
of Tab.~\ref{Table:1}. It turns out  that these ratios for different lattices 
and widths are clearly compatible with a common value, as required by 
universality (see last column of Tab.~\ref{Table:2}).  

\begin{table}[hbt]
\begin{center}
\begin{tabular}{|c|c|c|c|}
\hline
Lattice&$1/T$&$p_\ell$&$T_c/\sqrt{\sigma}$\\
\hline
SC site&7&0.3459514(12)&1.494(11)\\
\hline
BCC bond& 3&0.21113018(38)&1.497(10)\\
\hline
BCC bond& 4&0.20235168(59)&1.506(11)\\
\hline
SC bond&5&0.278102(5)  &1.480(12)\\
\hline
SC bond&6&0.272380(2)&1.492(13)\\
\hline
SC bond&7&0.268459(1)&1.500(13)\\
\hline
SC bond&8&0.265615(5)&1.504(14)\\
\hline
\end{tabular}
\end{center}

\caption{The critical $p_\ell$  for  different lattices at different 
temperatures and the corresponding universal ratio 
$T_c/\sqrt{\sigma}$ as obtained by combining Eq.s~(\ref{sscaling}) and (\ref{tscaling}). Errors in parenthesis affect the 
last digits.}
\label{Table:2}
\end{table}

In $(d+1)$-dimensional 
field theory at finite temperature there is a 
characteristic interplay between $d+1$ and $d$ critical behaviours. This is 
particularly evident in the present instance: the $p_\ell$ values are 
extracted through a  2-D percolation power law (\ref{tscal}) in order to 
take into account the finite size scaling tied to $L$. However the
 tower of $p_\ell$ values as a function of the width $\ell$ obeys a 
typical power law of 3-D percolation:        
\eq
p_\ell=p_c+\frac1{(T_c\,\ell)^{1/\nu}}~,
\label{tscaling}
\en
where $\nu$ indicates, as in all the other formulae of this paper, 
the thermal exponent of 3-D percolation, $p_c$ is the 
critical threshold as listed in Tab.~\ref{Table:1} and the amplitude $T_c$ 
depends on the kind of lattice. Figure~\ref{Figure:10} 
shows the four $p_\ell$ values of the bond SC  lattice as a function
of $\ell^{-1/\nu}$. A one-parameter fit to Eq.~(\ref{tscaling}) yields
\eq
T_c=4.45\pm0.02~~.
\label{tc}
\en
It should be noted parenthetically that, as the $p_\ell$'s are essentially 
two-di\-men\-sion\-al quantities, they can be evaluated with high precision. 
Comparison between Table~\ref{Table:1} and Table~\ref{Table:2}  shows that 
in some cases the level of 
precision of those  overcomes that of the best estimates 
of $p_c$. Moreover, with a modest additional computational effort, it would 
be possible to further improve their precision. 
Thus one could  perhaps envisage to apply systematically 
Eq.~(\ref{tscaling}) to improve the estimates of $p_c$ and/or $\nu$.
\[
\begin{array}{cc}
\refstepcounter{figure}
\label{Figure:9}
\epsfxsize=.471\textwidth
\epsfbox{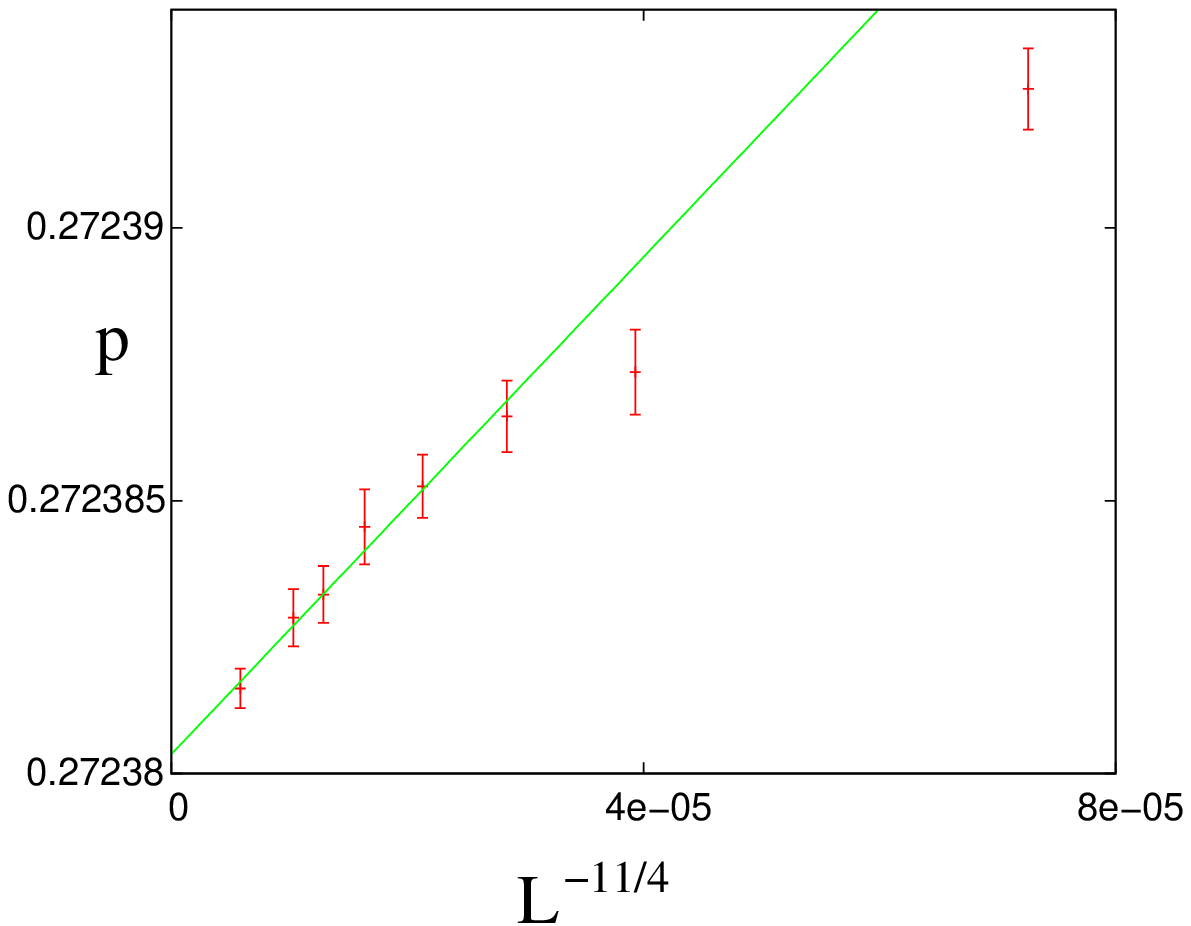}
&
\refstepcounter{figure}
\label{Figure:10}
\epsfxsize=.471\textwidth
\epsfbox{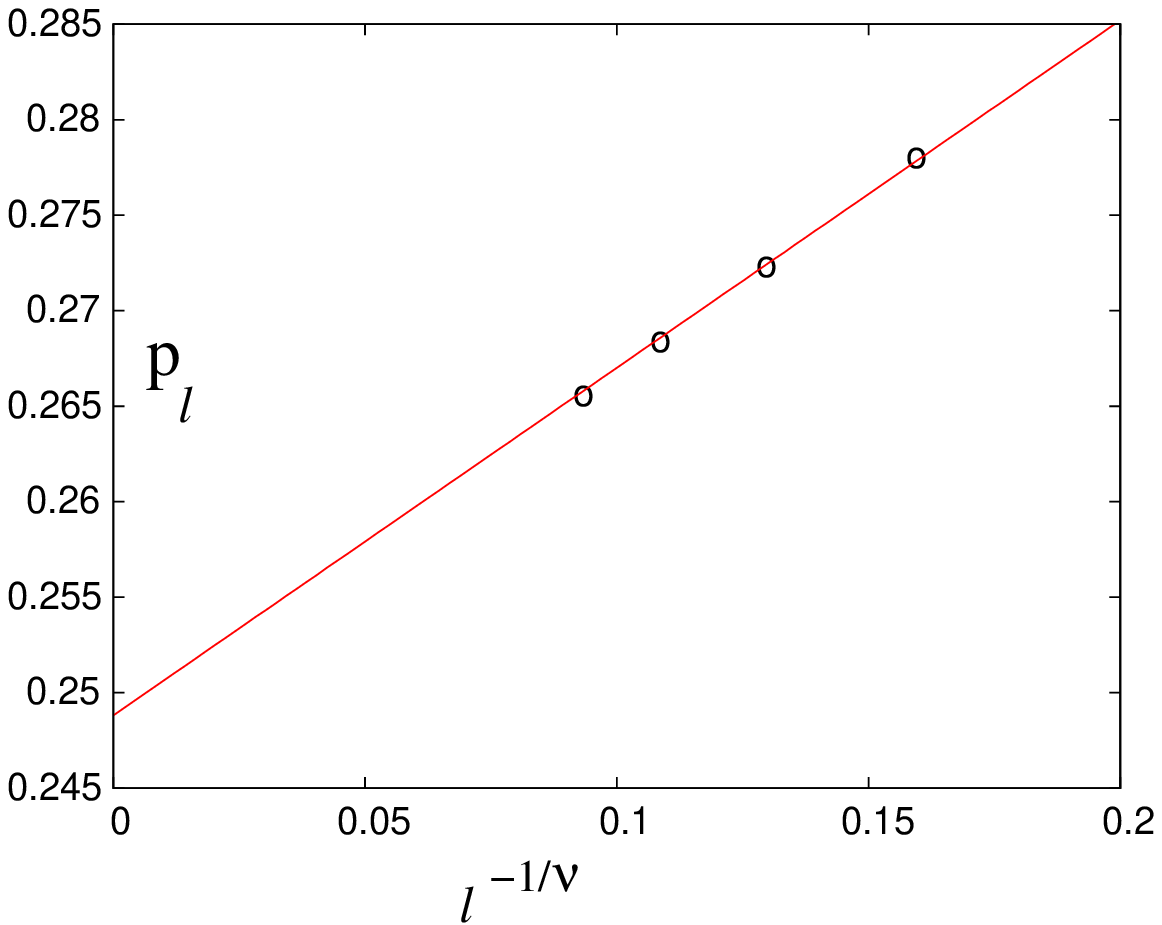}

\\
\parbox{.471\textwidth}{\small \raggedright
Figure~\ref{Figure:9}:
Finite-size scaling of the estimated $L\to\infty$ value of 
$p_\ell$ for bond percolation in a slice of SC lattice of size 
$L\times L\times 6$  with $32\leq L\leq80$. Each point
corresponds to about $10^8$ configurations. The solid line is a fit to 
Eq.~(\ref{tscal}).} &
\parbox{.471\textwidth}
{\small \raggedright Figure~\ref{Figure:10}:
Finite-size scaling of $p_\ell$ as a function of $\ell^{-1/\nu}$, where 
$\ell$ is the width of a SC lattice in bond percolation.The solid line is a 
one-parameter fit to Eq.~(\ref{tscaling}). The statistical errors are 
much smaller than the symbol size. }
\end{array}
\]

\resection{Universality class of deconfinement}
\label{universality}
The deconfined phase of any  $(d+1)$-dimensional 
gauge theory at finite 
temperature is characterised by the vanishing of the string 
tension $\sigma$. The interaction
between static sources (quarks) is described in terms of Polyakov operators. 
These are straight Wilson  loops wrapped around the short periodic  
direction $\ell$ (see Fig.~\ref{Figure:11}). The Polyakov-Polyakov correlator  
$\bra P(0)P(\vec{R})\ket$ of two parallel Polyakov operators 
only depends on their relative
positions in the $d$-dimensional sub-lattice. 
At the deconfining point it is expected to obey a power law dictated by the 
universality class of the transition. 

The critical behaviour of gauge theories at the deconfining temperature is
well described by the  Svetitsky-Yaffe (SY) conjecture \cite{sy} which can be 
formulated as follows. Suppose a $d+1$ dimensional gauge theory with 
gauge group $G$ has a second-order deconfinement transition at a certain 
temperature $T_c$; then its universality class is the same of the order-disorder transition of a  $d$-dimensional spin system with a global symmetry 
group coinciding  with the center of the gauge group. In particular, 
the Polyakov-Polyakov correlator corresponds to the spin-spin correlator of 
such a $d$-dimensional system. Therefore, at the critical point, it should 
decay as  
\eq
\bra P(0)P(\vec{R})\ket\,=\frac {\rm const.}{R^{d-2+\eta}}~~,
\label{PP}
\en
where $\eta$ is the magnetic exponent of the spin model.
The validity of this conjecture has been well 
established in a large number of numerical studies.

In the present case the above conjecture requires some generalisation, 
owing to the fact that not only the center, but the whole gauge 
group is trivial. Actually the SY conjecture is somehow related to the 
dimensional crossover in a layered lattice system \cite{cf}. 
The universality class of such a 
system, as it approaches a critical point, depends on the number of 
spatial directions which are going to infinity in the thermodynamic 
limit. This simple observation provides the basis for arguing  that the 
critical behaviour of percolation in a slab of finite thickness is well 
described by the 2-D percolation universality class. This is also supported 
by the fact that  finite size scaling of  threshold probability
$p_\ell$ is driven by the thermal exponent $\nu_2$, as Eq.~(\ref{tscal}) 
and Figure~\ref{Figure:9} show.

Such a conclusion is much less obvious, and more interesting, when considering 
the Polyakov-Polyakov correlator. This quantity can be evaluated using exactly 
the same method described in Section~\ref{cut} for the Wilson loops. One 
simply has to take into account that in the Wilson loop the surface 
$\Sigma$ (see Fig.~\ref{Figure:2}) used to evaluate  cluster wrapping 
is a rectangle, while in the latter case  is a cylinder bounded by the 
the two Polyakov lines. Periodic 
boundary  conditions put into play another difference: there are now four 
topologically different surfaces bounded by the two Polyakov lines (see 
Figure~\ref{Figure:12}), hence the correlator must be written as the sum 
of these four contributions. When the distance $R$ between  Polyakov 
lines is much smaller than  lattice size $L$ the main contribution comes from 
the top two surfaces of Figure~\ref{Figure:12}. 

One instance is reported in Figure~\ref{Figure:13}, where we plot the 
estimated  Polyakov-Polyakov correlator extracted from 
$10^4$ configurations in a $200\times200\times6$ lattice at the 
critical value of $p_6$ as determined in Tab.~\ref{Table:2}. Plotting  these 
data versus $R^{-\eta}$ with $\eta=\frac5{24}$ shows linear behaviour, as 
expected for a critical system which lies in the universality class 
of 2-D percolation.

\[
\begin{array}{cc}
\refstepcounter{figure}
\label{Figure:11}
\epsfxsize=.471\textwidth
\epsfbox{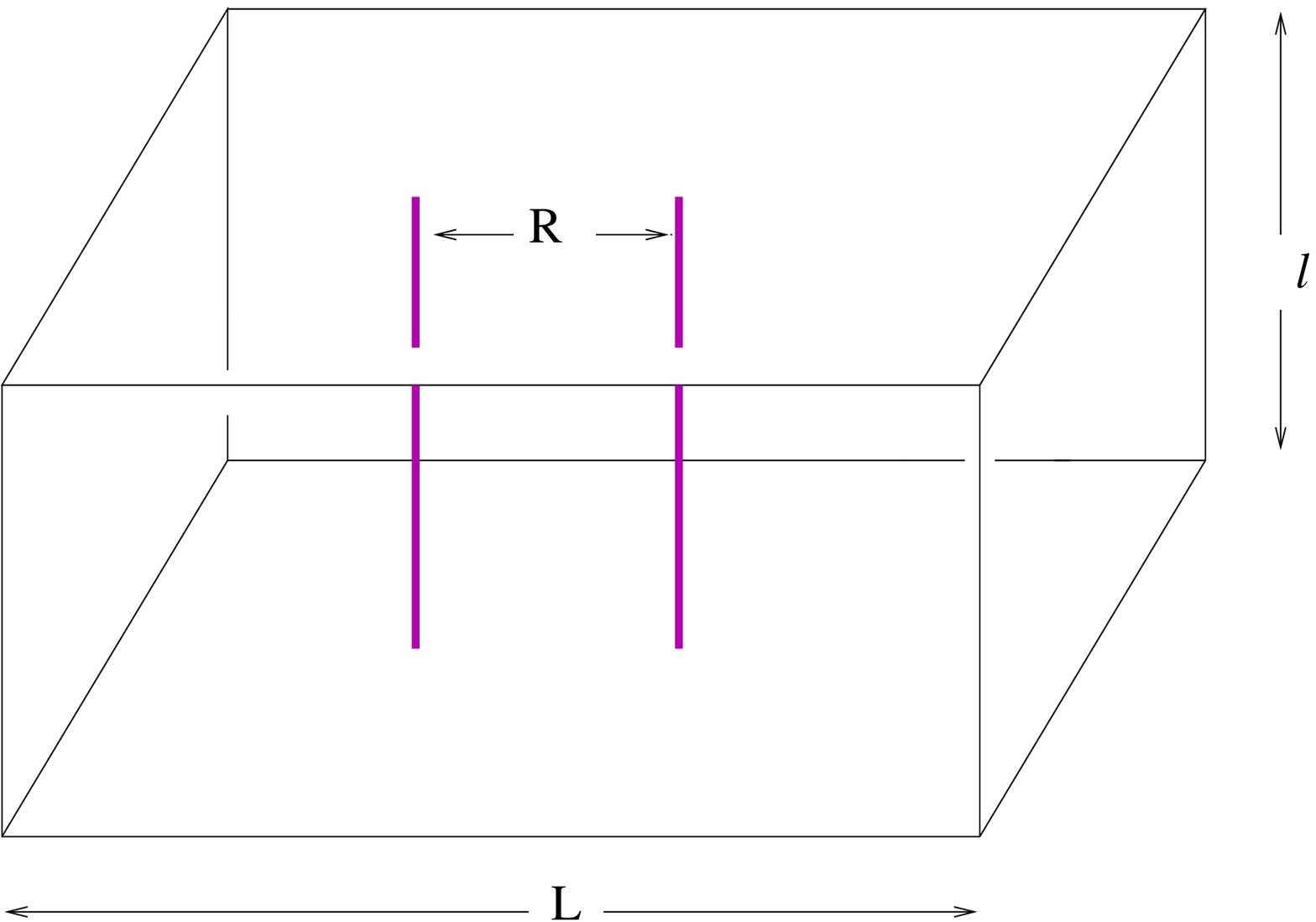}
&
\refstepcounter{figure}
\label{Figure:12}
\epsfxsize=.471\textwidth
\epsfbox{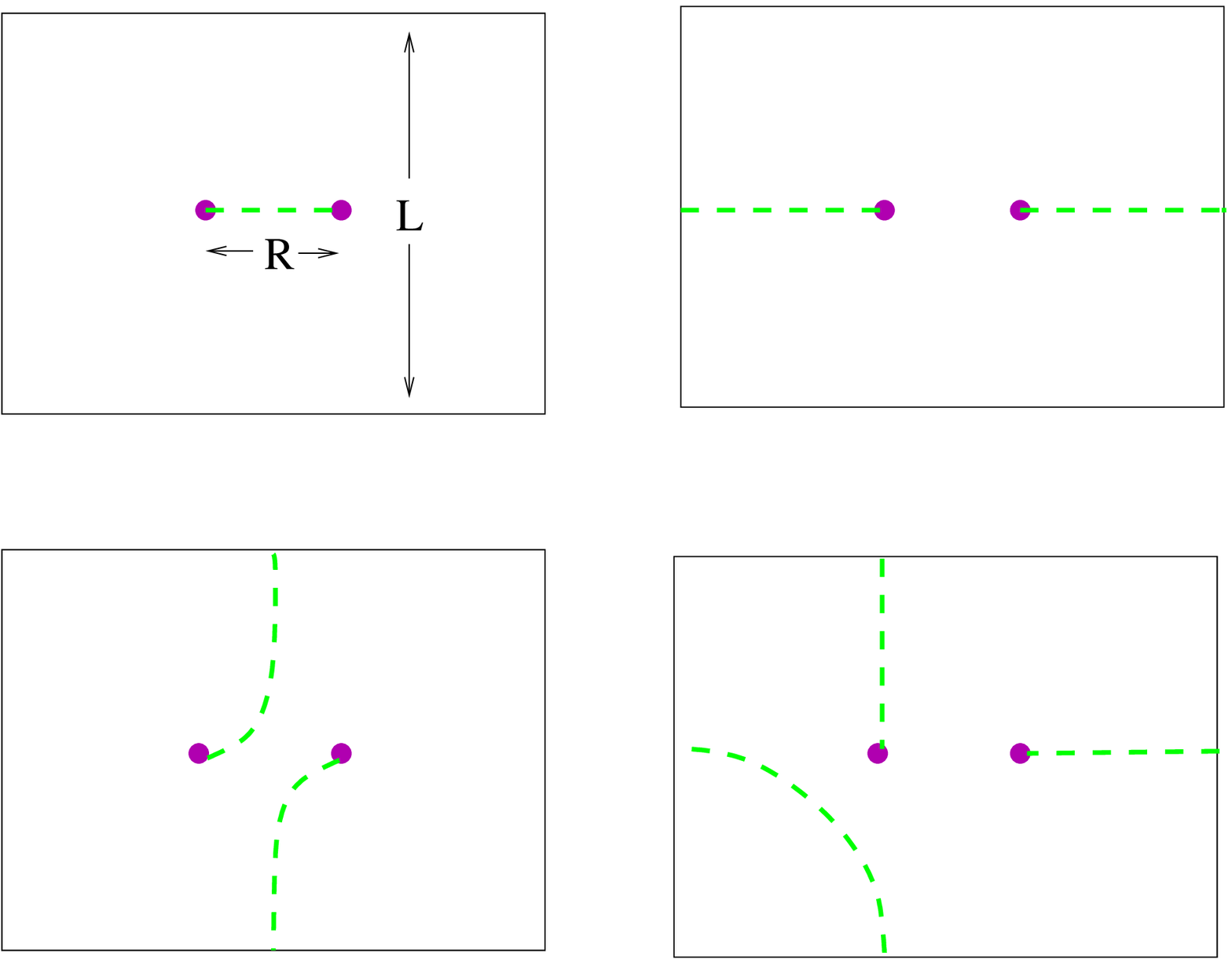}

\\
\parbox{.471\textwidth}{\small \raggedright
Figure~\ref{Figure:11}:
The slab geometry of the finite temperature setting. The two thick 
lines represent two  Polyakov loops wrapping in the periodic temperature 
direction.} &
\parbox{.471\textwidth}
{\small \raggedright Figure~\ref{Figure:12}:
The dashed lines represent the 1-D section of the four topologically
different surfaces bounded by the two Polyakov loops (solid circles)
in the case of periodic bc. }
\end{array}
\]

\begin{figure}[htb]
\begin{center}
\centering
\includegraphics[width=0.9\textwidth]{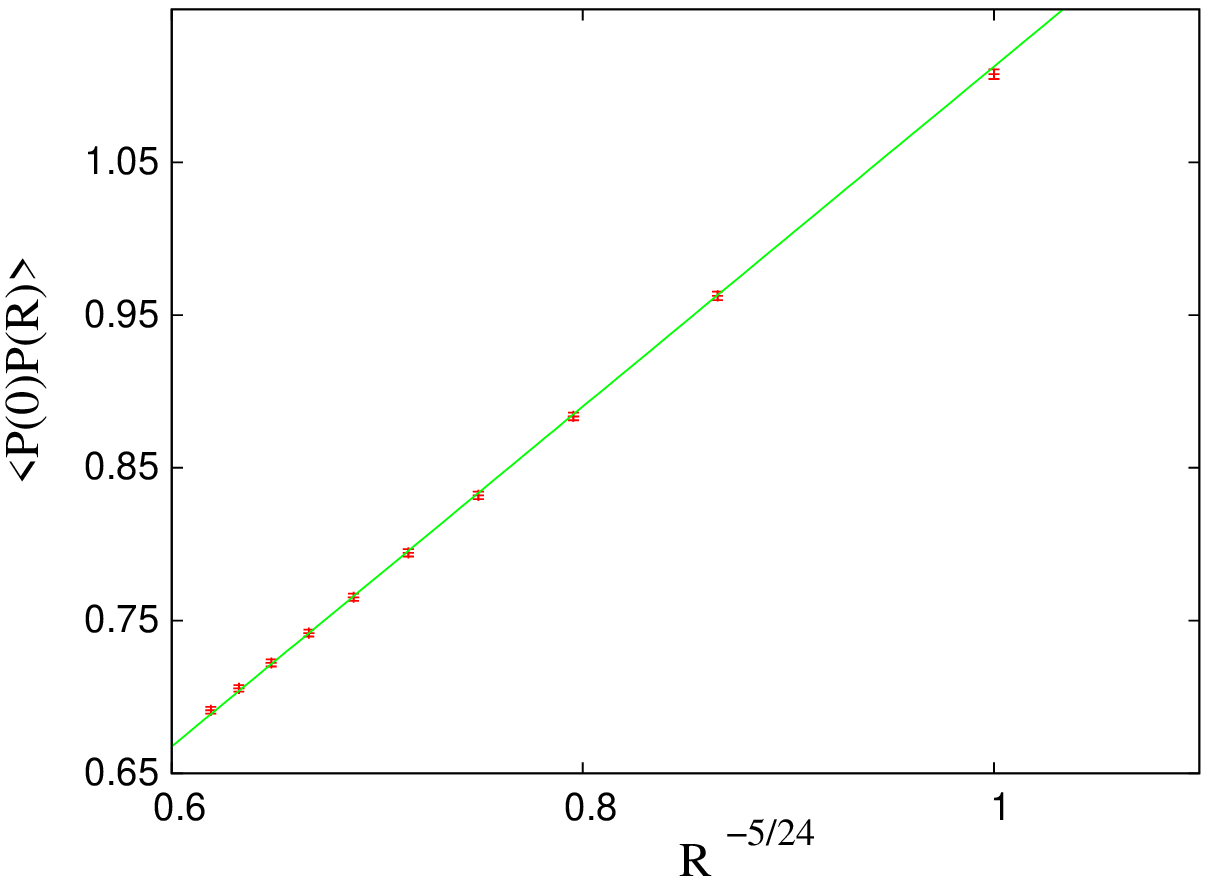} 
\caption{Polyakov-Polyakov correlator at  deconfing temperature 
as a function of the scaling variable $R^{-\eta}$, where $\eta=\frac5{24}$ is
the magnetic exponent of 2-D percolation.The lattice size is 
$200\times200\times6$. The solid line is a one-parameter fit to Eq.~(\ref{PP}) 
of the data with $R>2$. The resulting $\chi^2/d.o.f.$ is 0.46. }
\label{Figure:13}
\end{center}
\end{figure}

\resection{Conclusions}
\label{conclusion}
We have studied some consequences of a new point of view in three-dimensional 
random percolation which allows us to reinterpret it as a full-fledged 
gauge theory. 

A key difference between the conventional and the present 
approach is that instead of studying the universal scalings through the 
size distribution of  random clusters we  only consider their topological
entanglement with suitable closed paths. This suggests a reformulation in 
the language of gauge theory. In this context  a more  detailed 
description of 3-D percolation universality class emerges as 
a major issue: the gauge-percolation dictionary leads to  define 
 new physical quantities that can be used to extract new universal 
amplitude ratios. 

A typical example is the ratio $T_c^2/\sigma$ between
the (square of) deconfining temperature $T_c$ and the string tension $\sigma$. 
We evaluated it on seven different lattices (see Tab.~\ref{Table:2}) finding
excellent agreement with universality. Notice that neither $T_c$ nor 
$\sigma$ are truly foreign concepts of percolation theory: slab 
percolation has  always been a subject of intensive study. 
Strictly speaking, string tension  does not seem to 
have been considered previously in percolation studies, being a typical 
notion of gauge theory. There is however an intimately related quantity, 
the surface tension, which can also be defined 
in percolation \cite{harris}. The novelty introduced by the gauge theory 
interpretation is a previously unsuspected relationship between slab 
percolation and surface tension.

Another class of universal amplitude ratios came to us as a surprise. It 
turns out that the Wilson loop correlators receive contribution from a 
tower of physical states of increasing mass 
and spin, like in ordinary gauge theories. The ratios of their masses 
define universal quantities which further characterise  the universality class
of 3-D percolation.

Transcribing percolation into  gauge theory language has also some 
interesting consequences in the study of quark confinement mechanisms.  
The confinement generated by a random percolating cluster  is similar, but 
not identical, to that produced by an infinite network of center vortices. 
In particular the latter carry some conserved charge which regulates 
their mutual intersections, while the former do not 
carry any conserved  charge and intersect freely. This indicates that 
the intersection rules of center vortices do not play an important role 
in producing confinement.

One of the most surprising findings of the  present  approach is the 
observation  of some   universal shape effects in the 
vacuum expectation value of Wilson loops  which have been always ascribed to a 
different picture of confinement. This picture says that the 
flux of the gauge field generated by pair of quarks is squeezed by the
magnetic monopole condensate into a string-like structure which 
can vibrate freely. It turns out that  
these very vibrations
generate the universal effects mentioned above. Apparently, the two 
different pictures of confinement are different descriptions of the 
same physical phenomenon.

 It would be interesting to extend our percolation approach to a 
4-D gauge theory. This would require the study of plaquette 
percolation and the Wilson loops should measure their entanglement with 
closed surfaces.

\vskip1.0cm {\bf
Acknowledgements.} M.~P. acknowledges support received from 
Enterprise Ireland under the Basic Research Programme.

\end{document}